\documentclass[prd,preprint,superscriptaddress,showpacs,byrevtex]{revtex4}
\usepackage{bm}
\def\fsl#1{\setbox0=\hbox{$#1$}           
   \dimen0=\wd0                                 
   \setbox1=\hbox{/} \dimen1=\wd1               
   \ifdim\dimen0>\dimen1                        
      \rlap{\hbox to \dimen0{\hfil/\hfil}}      
      #1                                        
   \else                                        
      \rlap{\hbox to \dimen1{\hfil$#1$\hfil}}   
      /                                         
   \fi}                                         %
\newcommand{\be}{\begin{equation}}
\newcommand{\ee}{\end{equation}}
\newcommand{\bea}{\begin{eqnarray}}
\newcommand{\eea}{\end{eqnarray}}
\newcommand{\beq}{\begin{equation}}
\newcommand{\eeq}{\end{equation}}
\newcommand{\beqs}{\begin{eqnarray}}
\newcommand{\eeqs}{\end{eqnarray}}

\newcommand{\aslash}{A\hspace{-0.067in}\slash}

\begin{document}
\title{ Hadron Formation From Quark-Gluon Plasma Using Lattice QCD At Finite Temperature }
\author{Gouranga C Nayak }\thanks{E-Mail: nayakg138@gmail.com}
%
%
\date{\today}
\begin{abstract}
Recently we have reported the correct formulation of the lattice QCD method at the zero temperature to study the hadron formation from the quarks and gluons by incorporating the non-zero boundary surface term in QCD which arises due to the confinement of quarks and gluons inside the finite size hadron. In this paper we extend this to the finite temperature QCD and present the correct formulation of the lattice QCD method at the finite temperature to study the hadron formation from the quark-gluon plasma.
\end{abstract}
\maketitle
\pagestyle{plain}

\pagenumbering{arabic}

\section{introduction}

Just after the $10^{-12}$ seconds of the big bang our universe was filled with a hot and dense state of matter known as the quark-gluon plasma (QGP). The temperature of the quark-gluon plasma is $\ge$ 200 MeV. The QGP is the densest state matter besides the black hole. Hence it is important to recreate this early universe matter in the laboratory \cite{npc}.

The relativistic heavy-ion colliders (RHIC) at BNL and the large hadron colliders (LHC) at the CERN are the two experiments which search the quark-gluon plasma in the laboratory. The RHIC experiment collides Au-Au at the total center of mass energy $\sqrt{s}$ = 200$\times$ 197 GeV and the LHC experiment collides Pb-Pb at the total center of mass energy $\sqrt{s}_{NN}$ = 5.2 $\times $ 208 TeV. Since these huge energies are deposited in very small volumes just after the nuclear collisions there is no doubt that the required energy density to create the quark-gluon plasma is reached at RHIC and LHC.

The main difficulty is to detect the quark-gluon plasma because we have not directly experimentally observed quarks and gluons. The quarks and gluons are confined inside the hadron. For this reason the indirect signatures are proposed for the quark-gluon plasma detection at RHIC and LHC. The prominent hadronic signatures for the quark-gluon plasma detection at RHIC and LHC are 1) $j/\psi$ suppression, 2) strangeness enhancement and 3) jet quenching.

It is well known that a hadron is a composite particle consisting of quarks and gluons which are the fundamental particles of the nature. The quarks and gluons are confined inside the hadron by the strong force (or the color force) which is described by the quantum chromodynamics (QCD) \cite{ymf} which is a fundamental theory of the nature.

The renormalized QCD \cite{tvf} is well understood at the small distance where the coupling becomes small due to the asymptotic freedom \cite{gwf,pof}. Hence the small distance partonic scattering cross section at the high energy colliders is studied by using the perturbative QCD (pQCD). Note that since the quarks and gluons are not directly experimentally observed the short distance partonic scattering cross section calculated at the high energy colliders by using the pQCD cannot be measured experimentally. Using the factorization theorem in QCD \cite{fcf,fc1f,fc2f} this short distance partonic level cross section is folded with the (experimentally extracted) parton distribution function (PDF) and the fragmentation function (FF) to compute the hadronic cross section which is experimentally measured at the high energy colliders.

The hadron formation from the quarks and gluons is a long distance phenomenon in QCD where the renormalized coupling becomes large due to the asymptotic freedom. Hence the hadron formation from the quarks and gluons can not be studied by using the pQCD. The non-perturbative QCD is necessary to study the hadron formation from the quarks and gluons. However, the non-perturbative QCD is not solved analytically. This is because of the presence of the cubic and quartic gluon field terms in the QCD lagrangian density which makes it impossible to solve the path integration in QCD analytically (see section III for details). For this reason the path integration in QCD is performed numerically in the Euclidean time.

The lattice QCD employs the numerical method to perform the path integration in QCD in the Euclidean time. Lattice QCD computes the non-perturbative correlation functions of the partonic operators in QCD. These partonic operators are chosen in the lattice QCD method in such a way that they carry the same quantum numbers of the hadron (see section III for details). The lattice QCD method inserts a complete set of hadronic states in between the partonic operators in the non-perturbative correlation function to study the hadronic observable from the quarks and gluons. One of the crucial assumption in the lattice QCD method is that it operates the unphysical QCD Hamiltonian of all the quarks plus antiquarks plus gluons inside hadron on the physical energy eigenstate of the hadron to obtain the physical energy eigenvalue of the hadron which is not correct. This is because of the non-zero boundary surface term (non-zero energy flux) in QCD due to the confinement of quarks and gluons inside the finite size hadron \cite{nkff}.

Recently we have presented the correct formulation of the lattice QCD method at the zero temperature to study the hadron formation from the quarks and gluons \cite{nkcf} by incorporating the non-zero boundary surface term in QCD which arises due to the confinement of quarks and gluons inside the finite size hadron \cite{nkff}. In this paper we extend this to the finite temperature QCD and present the correct formulation of the lattice QCD method at the finite temperature to study the hadron formation from the quark-gluon plasma.

The paper is organized as follows. In section II we discuss the non-zero boundary surface term (the non-zero energy flux) in QCD due to the confinement of quarks
and gluons inside the finite size hadron. In section III we describe the correct formulation of the lattice QCD method at the zero temperature to study the hadron formation from the quarks and gluons by incorporating the non-zero boundary surface term in QCD which arises due to the confinement of quarks and gluons inside the finite size hadron. In section IV we extend this to the finite temperature QCD and present the correct formulation of the lattice QCD method at the finite temperature to study the hadron formation from the quark-gluon plasma. Section V contains conclusions.

\section{ Non-Zero Boundary Surface Term in QCD Due to Confinement of Quarks and Gluons Inside Finite Size Hadron}

The conservation of energy in physics is derived from the first principle by using the time translational invariance in the Noether's theorem. In QCD the additional complication arises because the quark and the gluon fields are not gauge invariant. Hence the combined gauge transformation plus the time translation must be implemented together to derive the gauge invariant Noether's theorem in QCD from the first principle.

From the gauge invariant Noether's theorem in QCD one obtains the continuity equation \cite{nkgnf}
\bea
\partial_\nu T^{\nu \lambda}_{q{\bar q}g}(x)=0
\label{tpf}
\eea
where the gauge invariant energy-momentum tensor density $T^{\mu \nu }_{q{\bar q}g}(x)$ of the quark plus antiquark plus gluon in QCD is given by
\bea
T^{\nu \lambda}_{q{\bar q}g}(x)={\bar \psi}_k(x) \gamma^\mu [\delta^{kl}i \partial^\lambda -i gT^d_{kl}A^{d \lambda}(x)]\psi_l(x)+F^{\nu \sigma d}(x)F_{\sigma }^{~\lambda d}(x)+\frac{g^{\nu \lambda}}{4} F^{\mu \sigma d}(x)F_{\mu \sigma}^d(x)+(antiquark)\nonumber \\
\label{tmf}
\eea
where $\psi_i(x)$ is the quark field with color index $i=1,2,3$, the $A_\mu^a(x)$ is the gluon field with Lorentz index $\mu=0,1,2,3$, the color index $a=1,...,8$ and the non-abelian gluon field tensor is given by
\bea
F_{\nu \sigma}^d(x) = \partial_\nu A_\sigma^d(x) - \partial_\sigma A_\nu^d(x)+gf^{dhs} A_\nu^h(x) A_\sigma^s(x).
\label{fmf}
\eea
From the gauge invariant Noether's theorem in QCD we find from eq. (\ref{tpf}) that the energy $E_{partons}^H(t)$ of all the quarks plus antiquarks plus gluons inside the hadron $H$ satisfies the equation
\bea
\frac{dE_{partons}^H(t)}{dt}=-<H|\sum_{q,{\bar q},g} \int d^3r \partial_j  T^{j 0}_{q{\bar q}g}(t,r)|H>
\label{etf}
\eea
where
\bea
E_{partons}^H(t)=<H|H_{partons}^H|H>=<H|\sum_{q,{\bar q},g} \int d^3r T^{0 0}_{q{\bar q}g}(t,r)|H>.
\label{etf1}
\eea
In eqs. (\ref{etf}) and (\ref{etf1}) the $|H>$ is the energy eigenstate of the hadron $H$ normalized to unity, $\sum_{q,{\bar q},g}$ represents the sum over all the quarks, antiquarks and gluons inside the hadron $H$ and $H_{partons}^H$ is the Hamiltonian of all the quarks plus antiquarks plus gluons inside the hadron $H$.

Since the boundary surface term in QCD is at the finite distance due to the finite size of the hadron one finds that the boundary surface term in QCD is non-zero due to confinement of quarks and gluons inside the finite size hadron irrespective of the form of the $r$ dependence of the gluon field $A_\mu^a(t,r)$ and the $r$ dependence of the quark field $\psi_i(t,r)$ \cite{nkff}. Hence one finds that \cite{nkff}
\bea
<H|\sum_{q,{\bar q},g} \int d^3r \partial_j  T^{j 0}_{q{\bar q}g}(t,r)|H> \neq 0.
\label{etf2}
\eea
From eqs. (\ref{etf}) and (\ref{etf2}) one finds that the energy $E_{partons}^H(t)$ of all the quarks plus antiquarks plus gluons inside the hadron is not conserved, {\it i. e.},
\bea
\frac{dE_{partons}^H(t)}{dt}\neq 0.
\label{etf3}
\eea
From eqs. (\ref{etf}) and (\ref{etf2}) one finds, however, that the energy $[E_{partons}^H(t)+E_{flux}^H(t)]$ is conserved, {\it i. e.},
\bea
\frac{d[E_{partons}^H(t)+E_{flux}^H(t)]}{dt}=0
\label{etf4}
\eea
where the non-zero energy energy flux is given by [see eqs. (\ref{etf}) and (\ref{etf2})]
\bea
\frac{dE_{flux}^H(t)}{dt}=<H|\sum_{q,{\bar q},g} \int d^3r \partial_j  T^{j 0}_{q{\bar q}g}(t,r)|H>\neq 0.
\label{etf5}
\eea

\section{Hadron Formation From Quarks and Gluons at Zero Temperature QCD Using Lattice QCD }

The non-perturbative correlation function of the partonic operator ${\hat {\cal O}}^H(x)$ for the hadron $H$ formation in QCD at zero temperature is given by \cite{mtf,abf}
\bea
&&<0|{\hat {\cal O}}^H(x') {\hat {\cal O}}^H(x'')|0>=\frac{1}{Z[0]}\int [dA] [d{\bar \psi}][d\psi]~{\hat {\cal O}}^H(x') {\hat {\cal O}}^H(x'')~{\rm det}[\frac{\delta \partial^\mu A_\mu^c}{\delta \omega^d}]\nonumber \\
&& e^{i\int d^4x [-\frac{1}{4}F_{\mu \lambda}^c(x)F^{\mu \lambda c}(x) -\frac{1}{2\alpha} [\partial^\mu A_\mu^c(x)]^2+{\bar \psi}_l(x)[\delta^{lk}(i{\not \partial} -m)+gT^c_{lk}\aslash^c(x)]\psi_k(x)]}
\label{pcf}
\eea
where $|0>$ is the non-perturbative QCD vacuum ({\it i. e.}, the ground state of the full QCD, not the pQCD vacuum), $\psi_i(x)$ is the quark field, $A_\mu^a(x)$ is the gluon field, $\alpha$ is the gauge fixing parameter and $F_{\lambda \sigma}^c(x)$ is given by eq. (\ref{fmf}). Note that in eq. (\ref{pcf}) there is no ghost field because we are directly working with the ghost determinant ${\rm det}[\frac{\delta \partial^\mu A_\mu^c}{\delta \omega^d}]$.

The partonic operator ${\hat {\cal O}}^H(x)$ carries the same quantum numbers of the hadron $H$. For example for the pion $\pi^+$ we have
\bea
{\hat {\cal O}}^{\pi^+}(t,r) =d^\dagger(t,r)\gamma_5 u(t,r)
\label{pif}
\eea
where $u(x)$ and $d(x)$ are the quark fields for the up and down quark. Similarly for the proton $P$ we have
\bea
{\hat {\cal O}}^P(t,r) ={u}(t,r) C\gamma_5 d(t,r)u(t,r)
\label{pf}
\eea
and for neutron $N$ we have
\bea
{\hat {\cal O}}^N(t,r) = d(t,r)C\gamma_5 u(t,r)d(t,r)
\label{nf}
\eea
where $C$ is the charge conjugation operator.

The time evolution of the partonic operator ${\hat {\cal O}}^H(t,r)$ in the Heisenberg representation is given by
\bea
{\hat {\cal O}}^H(t,r) = e^{-it H_{partons}^H} {\hat {\cal O}}^H(0,r) e^{it H_{partons}^H}
\label{tef}
\eea
where $H_{partons}^H$ is the QCD Hamiltonian of the partons. The energy $E_{partons}^H(t)$ of all the quarks plus antiquarks plus gluons inside the hadron $H$ is given by eq. (\ref{etf1}). From eq. (\ref{etf1}) we find
\bea
H_{partons}^H|H>=E_{partons}^H(t)|H>.
\label{etf1a}
\eea

Inserting a complete set of hadronic energy eigenstates
\bea
\sum_n |H_n><H_n|=1
\label{enef}
\eea
and then using eqs. (\ref{tef}) and (\ref{etf1a}) in (\ref{pcf}) we find in the Euclidean time
\bea
\sum_r <0|{\hat {\cal O}}^H(t,r) {\hat {\cal O}}^H(0)|0>=\sum_n |<0|{\hat {\cal O}}^H(0)|H_n>|^2~e^{-\int dt E_{partons,~n}^H(t)}
\label{efts}
\eea
where the $\int dt$ is an indefinite integration and from eq. (\ref{etf1a}) we have
\bea
H_{partons}^H|H_n>=E_{partons,~n}^H(t)|H_n>,~~~~~~~~~~~~|H_0>=|H>.
\label{etf1b}
\eea

Assuming that the contributions from all the higher energy level is neglected at the large time $t\rightarrow \infty$ we find from eq. (\ref{efts})
\bea
[\sum_r <0|{\hat {\cal O}}^H(t,r) {\hat {\cal O}}^H(0)|0>]_{t\rightarrow \infty}=|<0|{\hat {\cal O}}^H(0)|H>|^2~ e^{-\int dt E_{partons}^H(t)}
\label{efts1}
\eea
where $E_{partons}^H(t)$ is given by eq. (\ref{etf1}). From eq. (\ref{etf4}) we find that the energy $E^H$ of the hadron is given by
\bea
E^H=E_{partons}^H(t)+E_{flux}^H(t)
\label{ehf}
\eea
where the energy flux $E_{flux}^H(t)$ is given by eq. (\ref{etf5}).

Using eq. (\ref{ehf}) in (\ref{efts1}) we find
\bea
[\sum_r <0|{\hat {\cal O}}^H(t,r) {\hat {\cal O}}^H(0)|0>]_{t\rightarrow \infty}=|<0|{\hat {\cal O}}^H(0)|H>|^2~e^{-tE^H} e^{\int dt E_{flux}^H(t)}.
\label{efts2}
\eea
The energy flux as given by eq. (\ref{etf5}) can be calculated from the vacuum expectation in QCD by using the formula \cite{nkfx}
\bea
\frac{dE_{flux}^H(t)}{dt}=[\frac{<0| \sum_{r''}{\hat {\cal O}}^H(t'',r'') \sum_{q,{\bar q},g} \int d^3r \partial_j  T^{j 0}_{q{\bar q}g}(t,r){\hat {\cal O}}^H(0) |0>}{<0| \sum_{r''}{\hat {\cal O}}^H(t'',r'') {\hat {\cal O}}^H(0) |0>}]_{t''\rightarrow \infty}
\label{etfs3}
\eea
where the energy-momentum tensor density $T^{\nu \lambda}_{q{\bar q}g}(x)$ in QCD is given by eq. (\ref{tmf}).

Using eq. (\ref{etfs3}) in (\ref{efts2}) we find that for the hadron at rest
\bea
|<0|{\hat {\cal O}}^H(0)|H>|^2~e^{-tM^H}=\left[\frac{\sum_r <0|{\hat {\cal O}}^H(t,r) {\hat {\cal O}}^H(0)|0>}{e^{[\frac{<0| \sum_{r''}{\hat {\cal O}}^H(t'',r'') \sum_{q,{\bar q},g} \int d^4x \partial_j  T^{j 0}_{q{\bar q}g}(x){\hat {\cal O}}^H(0) |0>}{<0| \sum_{r''}{\hat {\cal O}}^H(t'',r'') {\hat {\cal O}}^H(0) |0>}]_{t''\rightarrow \infty}}}\right]_{t\rightarrow \infty}
\label{efts4}
\eea
where the $\int dt$ is an indefinite integration and $\int d^3x$ is definite integration in $\int d^4x=\int dt \int d^3x$.

Hence one finds that the hadronic decay matrix element $<0|{\hat {\cal O}}^H(0)|H>$ and the hadron mass $M^H$ can be found from the vacuum expectation value of the correlation functions in QCD at zero temperature as given by eq. (\ref{efts4}).

\section{Hadron Formation From Thermal Quark-Gluon Plasma Using Lattice QCD }

For the scalar field theory at the finite temperature $T$ the partition function $Z[J]$ is given by
\bea
Z[J]=\int [d\phi] e^{-\int_0^{\frac{1}{T}} dt \int d^3x [\partial_\mu \partial^\mu \phi(x)+V[\phi(x)]+J(x)\phi(x)]}={\rm Tr}~e^{-\frac{H}{T}}
\label{sftf}
\eea
where $t$ is the Euclidean time, $J(x)$ is the external current density, $V[\phi(x)]$ is the potential and the scalar field $\phi(x)$ satisfies the periodic boundary condition
\bea
\phi(t,{\vec x})=\phi(t+\frac{1}{T},{\vec x}).
\label{pbcf}
\eea
We use the Euclidean metric
\bea
g^{\mu \nu}=(-1,-1,-1,-1)
\label{emtf}
\eea
for the finite temperature field theory formulation in the Euclidean time.

Extending eq. (\ref{sftf}) to QCD we find that the generating functional at the finite temperature QCD is given by
\bea
&&Z[J,\eta,{\bar \eta}]=\frac{1}{Z[0]} {\rm Tr}~e^{-\frac{H}{T}}=\int [dA] [d{\bar \psi}][d\psi]\times {\rm det}[\frac{\delta \partial^\mu A_{\mu}^c}{\delta \omega^d}]\times {\rm exp}[-\int_0^{\frac{1}{T}}dt \int d^3x [-\frac{1}{4}F_{\mu \lambda }^c(x)F^{\mu \lambda c}(x)\nonumber \\
&&-\frac{1}{2\alpha} [\partial^\mu A_{\mu }^c(x)]^2+{\bar \psi}_{l}(x)[\delta^{lk}(i{\not \partial} -m)+gT^c_{lk}\aslash^c(x)]\psi_{k}(x)+J_\mu^a(x)A^{\mu a}(x)+{\bar \eta}_i(x) \psi_i(x)+{\bar \psi}_i(x) \eta_i(x)]] \nonumber \\
\label{pcfn1a}
\eea
where $J_\mu^a(x)$ is the source to the gluon field $A_\mu^a(x)$, the ${\bar \eta}_i(x)$ is the source to the quark field $\psi_i(x)$, the $F_{\mu \nu}^a(x)$ is given by eq. (\ref{fmf}) and the quark, gluon fields satisfy the periodic boundary condition
\bea
\psi_i(t,{\vec x})=\psi_i(t+\frac{1}{T},{\vec x}),~~~~~~~~~~~~~~~A_\mu^a(t,{\vec x})=A_\mu^a(t+\frac{1}{T},{\vec x}).
\label{pbcqf}
\eea

For the hadron $H$ formation from the thermal quark-gluon plasma we proceed as follows.
Similar to the QCD in vacuum case we choose the partonic operator ${\hat {\cal O}}^H(t,r)$ which form the hadron $H$ [see for example eqs. (\ref{pif}), (\ref{pf}) and (\ref{nf})] from the quark-gluon plasma.

The non-perturbative correlation function of the partonic operator ${\hat {\cal O}}^H(x)$ for the hadron $H$ formation from the quark-gluon plasma is given by
\bea
&&<in|e^{-t'H}{\hat {\cal O}}^H(t',r') {\hat {\cal O}}^H(0)|in>=\frac{1}{Z[0]} {\rm Tr}~e^{-\frac{H}{T}}~e^{-t'H}{\hat {\cal O}}^H(t',r') {\hat {\cal O}}^H(0)=\frac{1}{Z[0]}\int [dA] [d{\bar \psi}][d\psi]\nonumber \\
&&e^{-t'H}{\hat {\cal O}}^H(t',r') {\hat {\cal O}}^H(0)\times {\rm det}[\frac{\delta \partial^\mu A_{\mu}^c}{\delta \omega^d}]\times {\rm exp}[-\int_0^{\frac{1}{T}}dt \int d^3x [-\frac{1}{4}F_{\mu \lambda }^c(x)F^{\mu \lambda c}(x)
-\frac{1}{2\alpha} [\partial^\mu A_{\mu }^c(x)]^2\nonumber \\
&&+{\bar \psi}_{l}(x)[\delta^{lk}(i{\not \partial} -m)+gT^c_{lk}\aslash^c(x)]\psi_{k}(x)]]
\label{pcfn1}
\eea
where $|in>$ is the ground state of the full QCD at the finite temperature similar to the ground state $|0>$ of the full QCD at the zero temperature in eq. (\ref{pcf}).

Using the time evolution of the field as given by eq. (\ref{tef}) and then inserting complete set of hadronic states as given by eq. (\ref{enef}) we find from eq. (\ref{pcfn1}) in the Euclidean time that
\bea
&&\sum_{r'} <in|e^{-t'H}{\hat {\cal O}}^H(t',r') {\hat {\cal O}}^H(0)|in>=\sum_ne^{-\int dt'E_{partons,~n}^H(t')} |<H_n|{\hat {\cal O}}^H(0)|in>|^2
\label{pcfn5}
\eea
where the $\int dt'$ integral is indefinite integral and we have used the eq. (\ref{etf1b}) for the energy $E_{partons,~n}^H(t')$ of all the quarks plus antiquarks plus gluons inside the hadron $H$ in its n$th$ level.

Note that unlike QCD in vacuum where [see eq. (\ref{efts})]
\bea
e^{tH}|0>=0
\label{vczf}
\eea
we find that for the QGP at the finite temperature we have
\bea
e^{tH}|in>\neq 0.
\label{mczf}
\eea
Hence unlike the non-perturbative correlation function
\bea
C_2^{QCD}(t',r')=<0|{\hat {\cal O}}^H(t',r') {\hat {\cal O}}^H(0)|0>
\label{cfvf}
\eea
for the hadron formation from the quarks and gluons in QCD in vacuum in eq. (\ref{efts}) we have considered the non-perturbative correlation function
\bea
C_2^{QGP}(t',r')=<in|e^{-t'H}{\hat {\cal O}}^H(t',r') {\hat {\cal O}}^H(0)|in>
\label{cfmf}
\eea
for the hadron formation from the quarks and gluons from the quark-gluon plasma at the finite temperature in eq. (\ref{pcfn5}).

Note the difference of $e^{-t'H}$ between eqs. (\ref{cfvf}) and (\ref{cfmf}).

It should be mentioned here that the hadrons are in the confined phase of QCD whereas the quark-gluon plasma is in the deconfined phase of QCD. Hence the hadron can not be formed inside the quark-gluon plasma. When the quark-gluon plasma temperature $T$ decreases to the deconfinement phase transition temperature $T_c$ then the hadron is formed outside the quark-gluon plasma [{\it i. e.}, the hadron is formed in the vacuum] where $T>>>T_c$.

Since the hadron is formed in the vacuum [not inside the quark-gluon plasma medium] one finds that the energy $E_{partons,~n}^H(t')$ in eq. (\ref{pcfn5}) is the energy of all the quarks plus antiquarks plus gluons inside the hadron $H$ where the hadron $H$ is in the vacuum [see eqs. (\ref{etf1}) and (\ref{etf1b})].

Note that the Euclidean time $t$ goes to $\frac{1}{T}$ [see eq. (\ref{pcfn1})] which is the standard procedure to calculate the thermal average in the finite temperature quantum field theory. However, since the hadron $H$ is formed in the vacuum [not inside the quark-gluon plasma medium] one can take the Euclidean time $t'$ limit to infinity in the partonic operator ${\hat {\cal O}}^H(t',r')$ to form the hadron $H$ in vacuum. Because of this reason we cannot take $t$ to infinity but we can take take $t'$ to infinity.

Hence in the Euclidean time limit $t'\rightarrow \infty$ [similar to the QCD in vacuum case in eq. (\ref{efts1})] we can neglect the higher energy level contribution of the hadron to find from eq. (\ref{pcfn5}) that
\bea
\sum_{r'} [<in|e^{-t'H}{\hat {\cal O}}^H(t',r') {\hat {\cal O}}^H(0)|in>]_{t' \rightarrow \infty}=e^{-\int dt'E_{partons}^H(t')} |<H|{\hat {\cal O}}^H(0)|in>|^2.
\label{pcfn6}
\eea
As mentioned above since the hadron is formed in the vacuum [not inside the quark-gluon plasma medium] one finds that the energy $E_{partons}^H(t')$ of all the quarks plus antiquarks plus gluons inside the hadron $H$ is for the hadron $H$ in vacuum. Hence for the hadron $H$ in vacuum we can use eqs. (\ref{ehf}) and (\ref{etfs3}) from QCD in vacuum. Hence using eqs. (\ref{ehf}) and (\ref{etfs3}) in (\ref{pcfn6}) we find [similar to eq. (\ref{efts4})] that
\bea
|<in|{\hat {\cal O}}^H(0)|H>|^2=\left[\frac{\sum_{r'} <in|e^{-t'H}{\hat {\cal O}}^H(t',r') {\hat {\cal O}}^H(0)|in>}{e^{[\frac{<0| \sum_{r''}{\hat {\cal O}}^H(t'',r'') \sum_{q,{\bar q},g} \int d^4x' \partial_j  T^{j 0}_{q{\bar q}g}(x'){\hat {\cal O}}^H(0) |0>}{<0| \sum_{r''}{\hat {\cal O}}^H(t'',r'') {\hat {\cal O}}^H(0) |0>}]_{t''\rightarrow \infty}}}\right]_{t'\rightarrow \infty}\times ~e^{t'M^H}
\label{pcnf7}
\eea
where $\int dt'$ is an indefinite integration and $\int d^3x'$ is definite integration in $\int d^4x'=\int dt' \int d^3x'$.

Eq. (\ref{pcnf7}) describes the formation of hadron from the thermal quark-gluon plasma where the non-perturbative correlation function $<in|e^{-t'H}{\hat {\cal O}}^H(t',r') {\hat {\cal O}}^H(0)|in>$ is calculated by using the lattice QCD method at the finite temperature but the non-perturbative correlation functions $<0| {\hat {\cal O}}^H(t'',r'') \sum_{q,{\bar q},g} \int d^3x' \partial_j  T^{j 0}_{q{\bar q}g}(x'){\hat {\cal O}}^H(0) |0>$ and $<0|{\hat {\cal O}}^H(t'',r'') {\hat {\cal O}}^H(0) |0>$ are calculated by using the lattice QCD method at the zero temperature. Note that the hadron mass $M^H$ in eq. (\ref{pcnf7}) is in QCD in vacuum which can be calculated from the eq. (\ref{efts4}).

Since everything in the right hand side of eq. (\ref{pcnf7}) can be calculated by using the lattice QCD method at finite/zero temperature one finds that the probability $|<in|{\hat {\cal O}}^H(0)|H>|^2$ of the partons in the quark-gluon plasma at the finite temperature $T$ to form the hadron $H$ can be calculated by using the lattice QCD method by using eq. (\ref{pcnf7}).

\section{Conclusions}
Recently we have reported the correct formulation of the lattice QCD method at the zero temperature to study the hadron formation from the quarks and gluons by incorporating the non-zero boundary surface term in QCD which arises due to the confinement of quarks and gluons inside the finite size hadron. In this paper we have extended this to the finite temperature QCD and have presented the correct formulation of the lattice QCD method at the finite temperature to study the hadron formation from the quark-gluon plasma.

\end{document}